\def\rfr#1{eq. (\ref{#1})}

\def\derp#1#2{\rp{\partial{#1}}{\partial{#2}}}
\def\dert#1#2{\frac{{{d}}{#1}}{{{d}}{#2}}}

\def\virg#1{``#1''}

\def\eqi{\begin{equation}}
\def\eqf{\end{equation}}
\def\eqia{\begin{eqnarray}}
\def\eqfa{\end{eqnarray}}
\def\Om{\mathit{\Omega}}
\def\rp#1#2{{#1\over#2}} \def\lb#1{\label{#1}}

\def\kx{\hat{g}_x}
\def\ky{\hat{g}_y}
\def\kz{\hat{g}_z}
\def\bds#1{\boldsymbol{#1}}

\def\co{\cos\omega}
\def\so{\sin\omega}
\def\coo{\cos 2\omega}
\def\soo{\sin 2\omega}
\def\cO{\cos\Om}
\def\sO{\sin\Om}

\def\cI{\cos I}
\def\sI{\sin I}

\def\ton#1{\left(#1\right)}
\def\qua#1{\left[#1\right]}
\def\grf#1{\left\{#1\right\}}
\def\ang#1{\left\langle #1\right\rangle}

\documentclass[11pt]{article}
\usepackage{url}
\usepackage{amsmath,amsthm,amscd,amssymb}
\usepackage{latexsym,w-greek,wasysym}
\usepackage{graphicx,epsfig}
\usepackage{hyperref}
\allowdisplaybreaks[1]

\RequirePackage{color}

\begin{document}

\title{A uniform treatment of the orbital effects due to a violation of the Strong Equivalence Principle in the gravitational Stark-like limit}

\author{L. Iorio \\ Ministero dell'Istruzione, dell'Universit\`{a} e della Ricerca (M.I.U.R.)-Istruzione \\ Fellow of the Royal Astronomical Society (F.R.A.S.) \\
 International Institute for Theoretical Physics and
Advanced Mathematics Einstein-Galilei \\ Permanent address: Viale Unit$\grave{\rm a}$ di Italia 68
70125 Bari (BA), Italy \\ email: lorenzo.iorio@libero.it}

\maketitle

\begin{abstract}
\textcolor{black}{We consider a binary system made of self-gravitating bodies embedded in a constant and uniform external field $\bds g$}. We analytically work out several \textcolor{black}{orbital} effects \textcolor{black}{induced by a putative} violation of the Strong Equivalence Principle (SEP) \textcolor{black}{due to $\bds g$}. \textcolor{black}{In our calculation,} we do not \textcolor{black}{assume} $e\sim 0,$ \textcolor{black}{where $e$ is the binary's orbit eccentricity}. Moreover, \textcolor{black}{we do not a priori choose any specific preferred spatial orientation for the fixed direction of $\bds g$.}
Our results do not depend on any particular SEP-violating theoretical scheme. \textcolor{black}{They} can be applied to  general astronomical and astrophysical \textcolor{black}{binary systems immersed in an external constant and uniform polarizing field}.
%
\end{abstract}

\centerline

\centerline
\section{Introduction}
If bodies with non-negligible gravitational binding self-energies, like astronomical and astrophysical objects (planets and their natural satellites, main-sequence stars, white dwarfs, neutron stars), move with different accelerations in a given external field, the so-called Strong Equivalence Principle (SEP) would be violated. While the General Theory of Relativity (GTR) assumes the validity of SEP, it is generally violated in alternative theories of gravity; see \cite{Dam012,Freire012} and references therein.

Potential SEP violations in the Earth-Moon system freely falling in the external gravitational field of the Sun, theoretically predicted by Nordtvedt long ago \cite{nord1,nord2}, are currently searched for with the Lunar Laser Ranging (LLR) technique \cite{nord3,Tury07,LLR} in the framework of the Parametrized Post-Newtonian (PPN) formalism. There are projects to test SEP with the Martian moon Phobos and the Planetary Laser Ranging (PLR) technique \cite{Tury010} as well.

As shown by Damour and Sch\"{a}fer \cite{Dam91}, the acceleration due to SEP violation occurring for a two-body system in an external gravitational field $\bds g$ like, e.g., the Galactic field at the location of a binary pulsar, is, to leading order,
%
\eqi\bds A_{\textcolor{black}{g}}=\ton{\Delta_{\rm p}-\Delta_{\rm c}}\bds g,\lb{Asep}\eqf
where p and c denote the pulsar and its less compact companion; for a generic body with non-negligible gravitational self-energy\footnote{For a non-relativistic spherical body of uniform density, it can  be cast \cite{bind1,bind2} $\mathcal{E}_{\rm grav}=3M^2 G/5R$, where $G$ is the Newtonian constant of gravitation, $M$ is the body's mass, and $R$ is its radius. For  highly relativistic neutron stars, see, e.g., \cite{latt}.} $\mathcal{E}_{\rm grav}$, $\Delta$ accounts for the SEP-violating difference between the inertial mass $m_i$ and the gravitational mass $m_g$: at first post-Newtonian level, it is
\eqi
\Delta\doteq\rp{m_g}{m_i} -1 = \eta_1\varepsilon_{\rm grav},\ \varepsilon_{\rm grav}\doteq \rp{\mathcal{E}_{\rm grav}}{m_i c^2},
\eqf
where $c$ is the speed of light in vacuum. The external field $\bds g$ in \rfr{Asep} is assumed here to be constant and uniform, so that it induces a gravitational analog of the Stark effect \cite{Dam91}.
To maximize the size of potential SEP violations, one should look at compact objects like neutron stars: indeed, it is
\eqi\left|\varepsilon_{\rm grav}^{\leftmoon}\right|\sim 1.9\times 10^{-11},\ \left|\varepsilon_{\rm grav}^{\oplus}\right|\sim 4.2\times 10^{-10},\ \left|\varepsilon_{\rm grav}^{\odot}\right|\sim 1.3\times 10^{-6},\ \left|\varepsilon_{\rm grav}^{\rm NS}\right|\sim 10^{-1}.\eqf
The SEP-violating parameter $\eta_1$ can be parameterized in various ways depending on the theoretical framework adopted; see \cite{Dam012,Freire012} for  recent overviews. As far as the Earth-Moon system and the PPN framework is concerned, latest published bounds on the Nordtvedt parameter are
\begin{align}
\eta_1 & \equiv \eta_{\rm N}=(4.0\pm 4.3)\times 10^{-4}\ \cite{Tury07}, \\ \nonumber \\
\eta_1 & \equiv \eta_{\rm N}=(-0.6\pm 5.2)\times 10^{-4}\ \cite{LLR}.
\end{align}
Williams et al. in a recent analysis \cite{Williams012} released
\eqi
\eta_1\equiv \eta_{\rm N}=(-1.8\pm 2.9)\times 10^{-4}.
\eqf

In this paper, we deal with the orbital effects due to \rfr{Asep} in a uniform way by obtaining explicit and transparent analytical expressions valid for the main observable quantities routinely used in empirical studies. In Section \ref{kepelems} we work out the SEP\textcolor{black}{-violating} rates of change of the osculating Keplerian orbital elements, which are commonly  used in solar system and binaries studies. The SEP\textcolor{black}{-violating} shift of the projection of the binary's orbit onto the line-of-sight, which is the basic observable in pulsar timing, is treated in Section \ref{los}. In Section \ref{radvel} we calculate the SEP\textcolor{black}{-violating} perturbation of the radial velocity, which is one of the standard observable in spectroscopic studies of binaries: in systems suitable for testing SEP, the pulsar's companion is a white dwarf or a main sequence star for which radial velocity curves may be obtained as well. The primary-to-companion range and range-rate SEP\textcolor{black}{-violating} perturbations, potentially occurring in systems like the Earth-Moon one, are calculated in Section \ref{rrate}. Section \ref{discuss} is devoted to summarizing and discussing our results. \textcolor{black}{W}e will not resort to a-priori simplifying assumptions about the binary's orbital eccentricity. Moreover, we will not restrict ourselves to any specific spatial orientation for $\bds g$. Our results are, thus, quite general. They can be used for better understanding and interpreting  present and future SEP experiments, hopefully helping in designing new tests as well. Finally, we remark that our calculations are model-independent in the sense that they do not depend on any specific theoretical mechanism yielding SEP violations. Thus, they may be useful when different concurring SEP\textcolor{black}{-violating} scenarios are considered like, e.g., MOND and its External Field Effect \cite{Milg09,Blan011}, Galileon-based theories with the Vainshtein mechanism \cite{Gal012,Ior012}, and variations of fundamental coupling constants as well \cite{Dam011,Ior011,IorG}. Moreover, there are also other effects, like Lorentz symmetry violations \cite{BailLV}, which affect the same SEP\textcolor{black}{-violating} peculiar observables like the eccentricity \cite{IorLV}. \textcolor{black}{For a seminal paper on  Lorentz-violating orbital effects in binaries, see \cite{1992PhRvD..46.4128D}.}
\section{The long-term rates of change of the osculating Keplerian orbital elements}\lb{kepelems}
The SEP\textcolor{black}{-violating} acceleration of \rfr{Asep} is much smaller that the standard two-body Newtonian and Post-Newtonian ones; thus, its orbital effects can be treated perturbatively. It can be thought as due to the following perturbing potential
\eqi
U_{\textcolor{black}{g}}=\Xi\ {\bds g}{\bds\cdot}{\bds r},\lb{Potsep}
\eqf
where  \eqi\Xi\doteq \Delta_{\rm c}-\Delta_{\rm p}:\eqf
$\bds r$ is the relative position vector of the binary.

The Lagrange planetary equations \cite{BeFa} allow to work out all the  orbital effects caused by \rfr{Potsep} in an  effective way which encompasses just one integration.

Damour and Sch\"{a}fer \cite{Dam91} used a different formalism and an approach  requiring six independent\footnote{Averages of the time derivatives of the orbital energy, the orbital angular momentum and the Laplace-Runge-Lenz vector were taken in \cite{Dam91}. The orbital angular momentum and the Laplace-Runge-Lenz vector are perpendicular.} integrations. \textcolor{black}{The work by Damour and Sch\"{a}fer \cite{Dam91} is important also for the relevance for astrophysical tests found in the literature}.
Freire et al. \cite{Freire012} \textcolor{black}{recently used} the same formalism as Damour and Sch\"{a}fer \cite{Dam91}.

As a first step, the perturbing potential of \rfr{Potsep} must be evaluated onto a reference trajectory assumed as unperturbed with respect to the effect we are interested in. To this aim, we will adopt the standard Keplerian ellipse, although it would be possible, in principle, to use a Post-Newtonian orbit as well \cite{Calura1,Calura2}. We do not use such a relativistic unperturbed path \textcolor{black}{as a reference trajectory. Indeed, in addition to the usual 1PN pericenter precession  and the the  SEP\textcolor{black}{-violating}/0PN effects which will turn out to be \textit{per se} small,} it would only yield additional SEP\textcolor{black}{-violating}/1PN mixed terms.
\textcolor{black}{It must be stressed that, strictly speaking, the previous considerations-and the consequent calculations that we will show below-hold only in the case of a SEP violation due to an external polarizing field $\bds g$. In fact, a SEP-violating gravity theory is expected to modify the 1PN pericenter precession rate even if such an external field is absent; it is accounted for by the multiplicative factor $\mathcal{F}$ put by Damour and Sch\"{a}fer \cite{Dam91} in front of the standard Einsteinian 1PN pericenter precession, where $\mathcal{F}$ depends on the binary's masses in alternative theories. }
\textcolor{black}{Moving to our approach}, an average of \textcolor{black}{\rfr{Potsep}} over one full orbital period of the binary must be taken; by using the eccentric anomaly $E$ as fast variable of integration, the result is
\eqi
\ang{U_{\textcolor{black}{g}}} = -\rp{3\Xi ae g}{2} \grf{
\co\ton{\kx\cO + \ky\sO} + \so\qua{\kz\sI +\cI\ton{\ky\cO-\kx\sO}}},\lb{Ulss}
\eqf
where $g = \left|\bds g\right|, \bds{\hat{g}} = {\bds g}/{g},$
and $a,e,I,\omega,\Om$ are the semimajor axis, the eccentricity, the inclination, the argument of pericenter and the longitude of the ascending node of the binary orbit referred to a generic coordinate system.  As far as  binary pulsar systems are concerned, the reference $\{X,Y\}$ plane is the plane of the sky tangent to the celestial sphere at the location of the system considered. The inclination $I$ of the binary's orbital plane refers just to
the plane of the sky, so that the reference $Z$ axis is directed from the solar system's barycenter to the binary along the line-of-sight direction; the reference $Y$ axis  is usually directed towards the  North Celestial Pole, and the reference $X$ axis, from which the node $\Om$ is counted, is in the east direction. The unit vectors $\bds{I}_0,\bds{J}_0,\bds{K}_0$ of such a coordinate system can be expressed in terms of the Celestial right ascension (RA) $\alpha$ and declination (DEC) $\delta$ of the binary as \cite{Kop95,Kop96,Strat01}
\eqi
\bds{I}_0=\left\{
\begin{aligned}
& -\sin\alpha,  \\
& \cos\alpha,  \\
& 0,
\end{aligned}
\right.
\lb{vecx}
\eqf
\eqi
\bds{J}_0=\left\{
\begin{aligned}
& -\sin\delta\cos\alpha,  \\
& -\sin\delta\sin\alpha,  \\
& \cos\delta,
\end{aligned}
\right.
\lb{vecy}
\eqf
\eqi
\bds{K}_0=\left\{
\begin{aligned}
& \cos\delta\cos\alpha,  \\
& \cos\delta\sin\alpha,  \\
& \sin\delta.
\end{aligned}
\right.
\lb{vecz}\eqf
Instead, in studies pertaining our solar system the reference $\{X,Y\}$ plane is customarily chosen as coincident with the mean equator at J2000.0, so that the reference $X$ axis is directed towards the Vernal Equinox $\curlyvee$ at J2000.0 and the $Z$ axis points towards the North Celestial Pole.
In obtaining \rfr{Ulss} we have reasonably assumed that $\bds g$ does not vary over a full orbital revolution; moreover, we did not make any a priori simplifying assumption about the spatial orientation of $\bds g$ which, in general, will not be directed along any particular direction of the coordinate system adopted.
\textcolor{black}{Nonetheless,} the SEP\textcolor{black}{-violating}  field $\bds g$  can \textcolor{black}{reasonably}  be assumed to be approximately directed towards the Galactic Center (GC), whose Celestial coordinates are \cite{reid}
\begin{align}
\alpha_{\rm GC} & =17^{\rm h}45^{\rm m}37^{\rm s}.224=266.4051\ {\rm deg},\\ \nonumber \\
\delta_{\rm GC} & = -28^{\circ}56^{'}10^{''}.23=-28.936175\ {\rm deg}.
 \end{align}
 \textcolor{black}{Thus, for a Sun-planet pair in the Solar System it is
\eqi
\textcolor{black}{\bds{\hat{g}}\equiv}\bds{\hat{g}}_{\rm SS}=\left\{
\begin{aligned}
& \cos\delta_{\rm GC}\cos\alpha_{\rm GC}, \\
& \cos\delta_{\rm GC}\sin\alpha_{\rm GC} \\
& \sin\delta_{\rm GC}.
\end{aligned}
\right.
\lb{vecg}\eqf
}
%
%
%
%
%
%
%
%
%
%
%

The Lagrange planetary equations \cite{BeFa} and \rfr{Ulss} straightforwardly yield the following \textcolor{black}{long-term} rates of changes of the six standard osculating Keplerian orbital elements\footnote{$\mathcal{M}\doteq n_{\rm b}(t-t_0)$ is the mean anomaly, where $t_0$ is the time of passage at pericenter and $n_{\rm b}\doteq\sqrt{GM/a^3}$ is the Keplerian mean motion. \textcolor{black}{In general, SEP-violating theories induce modifications of the (effective) gravitational coupling parameter $G$ in the binary interactions (cfr. \cite{Dam91}); here we will neglect them.}}\textcolor{black}{, and of the longitude of the pericenter\footnote{\textcolor{black}{It is a \virg{dogleg} angle defined as $\varpi\doteq\Om + \omega$. It is used in some specific cases such as, e.g., analyses of planetary motions in our solar system.}} $\varpi$ as well.}
\begin{align}
\ang{\dert a t}  \lb{dadt2} & = 0, \\ \nonumber \\
\ang{\dert e t} \lb{dedt2}\nonumber & = -\rp{3\Xi g\sqrt{1-e^2}}{2n_{\rm b}a}\qua{\kz\sI\co + \cI\co\ton{\ky\cO-\kx\sO} -\right.\\ \nonumber \\
& - \left.\so\ton{\kx\cO +\ky\sO}  }, \\ \nonumber \\
\ang{\dert I t}  \lb{dIdt2}& = \rp{3\Xi g e\co\qua{\kz\cI + \sI\ton{\kx\sO-\ky\cO} }}{2n_{\rm b}a\sqrt{1-e^2}}, \\ \nonumber \\
\ang{\dert \Om t} \lb{dOdt2} & = \rp{3\Xi g e\csc I\so\qua{\kz\cI + \sI\ton{\kx\sO-\ky\cO}}}{2 n_{\rm b}a\sqrt{1-e^2}}, \\ \nonumber \\
\ang{\dert \omega t} \lb{dodt2} \nonumber & = \rp{3\Xi g}{2 en_{\rm b}a\sqrt{1-e^2}}\grf{\ton{-1+e^2}\co\ton{\kx\cO + \ky\sO} +\right.\\ \nonumber \\
& +\left. \so\qua{-\ky\cI\cO +\kz\ton{e^2\csc I -\sI} +\kx\cI\sO  }   }, \\ \nonumber \\
\ang{\dert \varpi t} \lb{dpdt2}\nonumber & = \rp{3\Xi g}{2en_{\rm b}a\sqrt{1-e^2}}\grf{\ton{1-e^2}\co\ton{\kx\cO + \ky\sO} +\right.\\ \nonumber \\
& + \left. \so\qua{\ton{\cI-e^2}\ton{\ky\cO-\kx\sO} + \kz\ton{1 - e^2 +\cI }\tan\ton{\rp{I}{2}}   } }, \\ \nonumber \\
\ang{\dert {\mathcal{M}} t}-n_{\rm b} \lb{dMdt2}\nonumber & = -\rp{3\Xi g\ton{1 + e^2}}{2en_{\rm b}a}\grf{\co\ton{\kx\cO+\ky\sO} +\right.\\ \nonumber \\
& + \left.\so\qua{\kz\sI +\cI\ton{\ky\cO-\kx\sO}} }.
\end{align}
We recall that, in obtaining the rate of the mean anomaly perturbatively, the mean motion $n_{\rm b}$ has to be  considered constant; thus, \rfr{dMdt2} is proportional to the rate of the time of pericenter passage. \textcolor{black}{About the pericenter precession, it must recalled that, in addition to the effect of \rfr{dpdt2} due to the external polarizing field $\bds g$, a further SEP-violating rate of change is, in principle, present. It is formally equal to the usual 1PN relativistic rate re-scaled by a multiplicative factor $\mathcal{F}$ which may depend on the masses of the binary in SEP-violating alternative theories of gravity \cite{Dam91}.}
In obtaining  \rfr{dadt2}-\rfr{dMdt2}, we did not make any a-priori simplifying assumption on the orbital configuration, i.e. we did not assume $e\rightarrow 0,\ I\rightarrow 0$. In the limit $e\rightarrow 0$, the orbital plane remains unchanged\textcolor{black}{, as found by Damour and Sch\"{a}fer \cite{Dam91},} since the long-term changes of both the inclination $I$ and the node $\Om$ vanish. Instead, the rate of change of the eccentricity $e$ remains substantially unaffected since it is not of order $\mathcal{O}(e)$. In the limit of small eccentricities \textcolor{black}{and small inclinations}, the mean longitude $l\doteq\varpi + \mathcal{M}$ is customarily used. From \rfr{dpdt2}-\rfr{dMdt2} it turns out that its rate is of order $\mathcal{O}(e)$:
\begin{align}
\ang{\dert l t}-n_{\rm b} \lb{dldt2}\nonumber & = -\rp{3\Xi e g}{4n_{\rm b} a} \grf{3\co\ton{\kx \cO + \ky\sO} + \so\qua{\ky\ton{2 + \cI}\cO-\right.\right. \\ \nonumber \\
&- \left.\left. \kx\ton{2+\cI}\sO + \kz\sI + 2\kz\tan\ton{\rp{I}{2}}}
}.
\end{align}
Thus, in the small eccentricity limit, all the osculating Keplerian orbital elements remains unaffected by a SEP violation, apart from the eccentricity.
Note that \rfr{dadt2}-\rfr{dMdt2} and \rfr{dldt2} are all proportional to $n_{\rm b}^{-1}a^{-1}=a^{1/2}$, so that they are larger for detached binaries.
For large eccentricities the node and the inclination variations can, in principle, be used as well as further observables. Interestingly, \rfr{dIdt2}-\rfr{dOdt2} show that their ratio is independent of both $g$ and $\bds{\hat{g}}$; moreover, it depends neither on $a$ nor on $e$, nor on $\Om$ amounting to
\eqi \rp{\ang{\dot\Om}}{\ang{\dot I}}=\csc I\tan\omega.\lb{rappor}\eqf
For systems \textcolor{black}{exhibiting large inclinations}, \rfr{rappor} may have relevant observational implications,
%
%
%
%
%
%
%
\textcolor{black}{provided that both the inclination and the node are accessible to observations}. Also the ratio of \rfr{dIdt2} and \rfr{dedt2} could be considered \cite{Freire012}; however, if on the one hand \textcolor{black}{the ratio of the rates of the inclination and the eccentricity} does not depend on $g$ and $a$, on the other hand it depends on $\bds{\hat{g}}$ and on $e,I,\Om,\omega$ as well, contrary to \rfr{rappor} \textcolor{black}{which is simpler}.

Damour and Sch\"{a}fer \cite{Dam91} define the following vector directed along the external field\footnote{Note that \textcolor{black}{the  quantity $f$ has the physical dimensions of  reciprocal time, i.e.} $[f] = {\rm T}^{-1}$.}
\eqi \bds f = \rp{3\bds A}{2n_{\rm b}a}.\eqf
 Then, they use the unit vector $\bds{\hat{a}}$ directed along the semimajor axis towards the pericenter, the unit vector $\bds{\hat{k}}$ directed along the orbital angular momentum, and $\bds{\hat{b}}=\bds{\hat{k}}\bds\times\bds{\hat{a}}$.
 Their expression for the averaged rate of change of the eccentricity is \cite{Dam91}
 \eqi \ang{\dert e t} = \sqrt{1-e^2}\ton{\bds{\hat{b}}\bds\cdot\bds f}.\lb{dedtDAM}\eqf
 A direct comparison of \rfr{dedtDAM} with our \rfr{dedt2} can actually be done by expressing the unit vectors $\bds{\hat{k}}$ and $\bds{\hat{a}}$ in terms of the Keplerian orbital elements as
\eqi
\bds{\hat{k}} = \left\{
\begin{aligned}
& \sI\sO,  \\
& -\sI\cO,  \\
& \cI.
\end{aligned}
\right.
\lb{veck}\eqf
and
\eqi
\bds{\hat{a}}=\left\{
\begin{aligned}
& \cO \co - \cI\sO\so,  \\
& \sO \co + \cI\cO\so,  \\
& \sI\so.
\end{aligned}
\right.
\lb{veca}\eqf
By using \rfr{veck}-\rfr{veca}, it turns out that \rfr{dedtDAM} coincides with our \rfr{dedt2}.

As far as the inclination $I$ is concerned, Freire et al. \cite{Freire012}, by adopting the same formalism of Damour and Sch\"{a}fer \cite{Dam91}, obtain
\eqi
\dert\cI t = \rp{e}{\sqrt{1-e^2}}\ton{\bds{\hat{K}}_0\bds\cdot\bds{\hat{b}}}\ton{\bds{\hat{k}}\bds\cdot\bds{\hat{f}}},\lb{didtDAM}
\eqf
where $\bds{\hat{K}}_0$ is the unit vector of the line-of-sight. Also in this case, with the aid of \rfr{veck}-\rfr{veca} it can be shown that \rfr{didtDAM} agrees with our \rfr{dIdt2}.

Neither Damour and Sch\"{a}fer \cite{Dam91} nor Freire et al. \cite{Freire012} explicitly considered the SEP\textcolor{black}{-violating} rates of change of the node (\rfr{dOdt2}), the pericenter (\rfr{dodt2}-\rfr{dpdt2}) and the time of pericenter passage (\rfr{dMdt2}).

In binaries hosting a radiopulsar, the rate of change of the inclination can be expressed in terms of the rate of change of the pulsar's projected semimajor axis $x_{\rm p}\doteq a_{\rm p}\sI/c,$ where $c$ is the speed of light in vacuum, as $\dot I=\left(\dot x_{\rm p}/x_{\rm p}\right)\tan I$.
About the measurability of the node $\Om$ in wide binaries hosting a radiopulsar, it could be determined only if they are close enough to the Earth. Indeed, in
this case the orbital motion of the Earth changes the apparent inclination angle $I$ of the
pulsar orbit on the sky, an effect known as the annual-orbital parallax \cite{Kop95}. It causes
a periodic change of the projected semi-major axis. There is also a second contribution due to the transverse motion in the plane of the sky \cite{Kop96}, yielding a secular variation of the projected semi-major axis. By including both these effects in the model of the
pulse arrival times, $\Om$ can be determined, as in
the case of PSR J0437-4715 \cite{Strat01}, located at only 140 pc from us. Interestingly, such an approach may also be used in optical
spectral observations of binary stars possessing a sufficiently well determined  radial velocity
curve \cite{Kop96}. As far as exoplanets are concerned, spectroscopic variations of the hosting star during the
transits can, in principle, be used to measure $\Om$ \cite{Quel00}. According to  Fluri and Berdyugina \cite{Fluri010},  the analysis of the polarization of the light scattered by the planetary
atmospheres may allow to determine, among other things, $\Om$ as well.

The SEP\textcolor{black}{-violating} orbital rates of \rfr{dadt2}-\rfr{dldt2}
can be defined as long-term effects since they were obtained by taking averages over one full orbital period $P_{\rm b}$ of the binary system. It must be noted that, strictly speaking, they cannot be considered as secular rates. Indeed, in calculating them, it was assumed that the external field $\bds g$ was constant over $P_{\rm b}$. The extent to which such an approximation can be considered valid depends on the specific system considered. It is certainly true for binary pulsar systems, for which timing data span some decades at most, when $\bds g$ is due to the Galaxy: in this case, $\bds g$ can reasonably be considered as constant over any foreseeable data analysis based on existing (and future) pulsar timing records because of the extremely low variation of the Galactic field at the pulsar's location during such time spans. If, instead, $\bds g$ is due to a remote third body with a small orbital frequency $n_{\rm b}^{'}$ with respect to $n_{\rm b}$, a relatively low modulation may be introduced by a slowly varying ${\bds g}(t)$. A quite different situation occurs for systems like the Earth-Moon one whose external field $\bds g$ is due to a relatively close body like the Sun having an orbital frequency $n_{\rm b}^{'}$  comparable to $n_{\rm b}$. In this case, $\bds g$ cannot be considered as constant, and a second integration has to be performed over the orbital period $P^{'}_{\rm b}$ of the third body.

In \rfr{dadt2}-\rfr{dldt2} the Keplerian orbital elements $I,\Om,\omega$ determining the orientation of the binary orbit in space are present. They are fixed only in the two-body Keplerian problem for two pointlike masses.
In any realistic situation, they are slowly varying with frequencies typically far smaller than  $n_{\rm b}$. It is true even if no other bodies are part of the system because of general relativity itself causing the well known 1PN gravitoelectric and gravitomagnetic precessions.

Thus, even if $\bds g$ can be considered constant, in general it is not so for $I,\Om,\omega$. Again, the peculiarities of the specific binaries at hand may greatly reduce the variability of them. For example, the more the system's orbit is large, the weaker are the effects of general relativity.

Our results of \rfr{dadt2}-\rfr{dldt2} are quite general, and can be applied to any gravitationally bound two-body system immersed in a (weak) external field: they are valid for any orbital geometry of the binary system and for a quite generic spatial orientation of $\bds g$ in a given coordinate system.
\section{The line-of-sight perturbation $\Delta\rho$}\lb{los}
In timing of binary systems typically hosting radiopulsars, the basic observable quantity is the projection $\rho$ of the pulsar's barycentric orbit along the line-of-sight \cite{Dam91} \textcolor{black}{since its variations $\Delta\rho$ are straightforwardly related to the timing via $\Delta\tau=\Delta\rho/c$}. In view of the fact that $\bds{\hat{\rho}}=\bds{\hat{Z}}$, one has to consider  the $Z$ component of the pulsar's barycentric position vector $\bds r$
\eqi Z = r\sI\sin\ton{\omega + \nu},\lb{zet}\eqf where $\nu$ is the true anomaly. The SEP\textcolor{black}{-violating} perturbation of $\rho$ due to \rfr{Asep} can be straightforwardly obtained in the following way:
\eqi \ang{\Delta\rho}=\int_0^{P_{\rm b}}\ton{\dert Z t}dt =\int_0^{2\pi}\qua{\derp Z{\textcolor{black}{E}}\dert {\textcolor{black}{E}}{\mathcal{M}}\dert{\mathcal{M}}t + \sum_{\kappa}\derp Z\kappa\dert\kappa t}\ton{\dert t{\textcolor{black}{E}}}d\textcolor{black}{E},\ \kappa=a,e,I,\omega,\lb{integrale}\eqf
where
\textcolor{black}{
\begin{align}
r \lb{raggius}& = a\ton{1 - e\cos E}, \\ \nonumber \\
\cos\nu & = \rp{\cos E - e}{1 - e \cos E}, \\ \nonumber \\
\sin\nu & = \rp{\sqrt{1 - e^2}\sin E}{1 - e \cos E}, \\ \nonumber \\
\dert E{\mathcal{M}} \lb{dEdM}& = \rp{1}{1 - e \cos E}, \\ \nonumber \\
\dert t E & = \rp{1 - e \cos E}{n_{\rm b}},
\end{align}
}
and $d\kappa/d t$ are the instantaneous rates of change of the Keplerian orbital elements computed onto the unperturbed Keplerian orbital ellipse according to, e.g., the right-hand-sides of the Gauss perturbative equations\footnote{In principle, $d{\mathcal{M}}/d t$ includes also $n_{\rm b}$, which yields the purely Keplerian part $\ang{\Delta\rho_{\rm Kep}}$ of the shift of the line-of-sight component of the orbit. It vanishes.} \cite{BeFa}.

\textcolor{black}{We have
\begin{align}
\ang{\Delta\rho} \lb{rhosep} \nonumber & = -\rp{\pi g\Xi\sqrt{1-e^2}\sI}{2e^4 n^2_{\rm b}}\grf{\ton{4 - 6 e^2 + 3 e^4}\kz\sI\soo + \right. \\ \nonumber \\
& + \left. \ton{4 - 6 e^2 + 3 e^4}\cI\soo\ton{\ky\cO - \kx\sO} + \right. \\ \nonumber \\
\nonumber & + \left. \qua{-e^4 + \ton{4 - 6 e^2 + 3 e^4}\coo }\ton{\kx\cO + \ky\sO}    }.
\end{align}
No approximations in $\bds{\hat{g}}, e$ and $I$ were used in deriving \rfr{rhosep}. A purely formal singularity appears in \rfr{rhosep} for $e\rightarrow 0$: actually, it is unphysical since it is cured by using properly chosen non-singular orbital elements (see, e.g., \cite{brou}).}

%
%
%
%
%
%
\section{The radial velocity perturbation $\Delta\dot\rho$}\lb{radvel}
 A standard observable in spectroscopic studies of binaries is the radial velocity. Up to the velocity of the binary's center of mass $V_0$, it is, by definition, $d\rho/dt$, i.e. the temporal rate of change of
the line-of-sight projection  $\bds r$ of the barycentric orbit of the binary's component whose light curve is available.

The unperturbed, Keplerian expression for $\dot\rho$ can straightforwardly be obtained from \rfr{zet} and \rfr{dfdM} as
\eqi \dot\rho=\derp Z \nu \dert \nu{\mathcal{M}}n_{\rm b}=\rp{n_{\rm b}a\ \textcolor{black}{\sI}}{\sqrt{1-e^2}}\qua{e\co + \cos\ton{\omega+\nu}}.\lb{vrad}\eqf
The SEP\textcolor{black}{-violating} perturbation $\ang{\dot\rho_{\rm SEP}}$ on $\dot\rho$ can be worked out by  replacing $Z$  \textcolor{black}{with \rfr{vrad} in \rfr{integrale}. For computational purposes, it turns out more convenient to use the true anomaly $\nu$ as fast variable of integration instead of the eccentric anomaly $E$.}

\textcolor{black}{ By using
\begin{align}
r \lb{raggius2} &= \rp{a\ton{1-e^2}}{1 + e\cos \nu}, \\ \nonumber \\
\dert \nu{\mathcal{M}} \lb{dfdM} & = \ton{\rp{a}{r}}^2\sqrt{1-e^2}, \\ \nonumber \\
\dert t \nu  & = \rp{\ton{1-e^2}^{3/2}}{n_{\rm b}\ton{1+e\cos \nu}^2},
\end{align}
 one finally gets
\begin{align}
\ang{\Delta\dot\rho} \lb{vrrr} \nonumber & =-\rp{9\pi g \Xi\sqrt{1-e^2}\sI}{e^2 n_{\rm b}}\grf{\kz\sI\coo + \right. \\ \nonumber \\
& + \left. \cI\coo\ton{\ky\cO - \kx\sO} -\soo\ton{\kx\cO + \ky\sO}
}.
\end{align}
Also in this case, the use of suitably defined non-singular orbital elements cures the formal singularity in \rfr{vrrr} for $e\rightarrow 0$.}
%
%
%
%
%
%
\section{The range and range-rate perturbations $\Delta r$ and $\Delta \dot r$}\lb{rrate}
By proceeding as in the previous Sections, it is possible to work out the SEP\textcolor{black}{-violating} range and range-rate perturbations $\Delta r$ and $\Delta \dot r$, respectively.
\textcolor{black}{More precisely, we used \rfr{integrale} in which we replaced $Z$ with \rfr{raggius2} for $\Delta r$, and with $dr/dt=n_{\rm b}ae\sin \nu/\sqrt{1-e^2}$ for $\Delta\dot r$.}
\textcolor{black}{Generally speaking,} such observable quantities are routinely measured in, e.g., the Earth-Moon LLR experiment\textcolor{black}{\footnote{\textcolor{black}{If the polarizing external field is due to the Sun, our calculation are not strictly applicable to the Earth-Moon system since the temporal variations of $\bds g$ may not be neglected over the characteristic timescales of such a binary.}}} and in several spacecraft-based interplanetary missions. The following calculations refer to the vector $\bds r$ connecting the centers of mass of the two bodies constituting the binary system; thus, they are valid when the distances of the ranging devices (laser retroreflectors, transponders on probes, etc.) from the centers of mass of the bodies are negligible with respect to $r$.

We obtain
\begin{align}
\ang{\Delta  r} \lb{range}\nonumber & = \rp{3\pi \Xi\sqrt{1-e^2}}{n_{\rm b}^2}\left[\kz\co\sI+\cI\co\left(\ky\cO-\kx\sO\right)-\right.\\ \nonumber \\
 & - \left.\so\left(\kx\cO+\ky\sO\right)\right], \\ \nonumber \\
\ang{\Delta \dot r} \lb{rangerate}\nonumber & =  \rp{3\pi \Xi\left[1+e\left(2-e\right)\right]}{\left(1-e^2\right)n_{\rm b}}\left\{
 \co\left(\kx\cO+\ky\sO\right) +\right. \\ \nonumber \\
 & + \left.\so\left[\kz\sI + \cI\left(\ky\cO-\kx\sO\right)\right]\right\}.
\end{align}
No approximations in $e$ were used in deriving \rfr{range} and \rfr{rangerate}.
\section{Summary and conclusions}\lb{discuss}
We worked out \textcolor{black}{the first time derivatives of some orbital effects} induced by a Stark-like SEP violation in a binary system made of two self-gravitating bodies immersed in an external polarizing field $\bds g$, assumed constant and uniform with respect to the characteristic temporal and spatial scales of the binary.

We provided the reader with full analytical expressions for the SEP\textcolor{black}{-violating} rates of change of all the six osculating Keplerian orbital elements (Section \ref{kepelems}), for the projection of the binary's orbit along the line-of-sight (Section \ref{los}) and its time derivative (Section \ref{radvel}), and for the range and range-rate (Section \ref{rrate}). We did not make any a-priori assumption about the orientation of $\bds g$ in space. We did not make simplifying assumptions about the orbital geometry of the binary.
%
%
%

It is important to have at disposal explicit expressions of the different relevant SEP\textcolor{black}{-violating} effects  since each one has a peculiar temporal pattern which may be helpful in separating it from other possible competing signatures. For example, PSR B1620-26 \cite{psr1,psr2} is a  pulsar-white dwarf binary  in a relatively wide orbit ($P_{\rm b}\textcolor{black}{= 16540653(6)\ {\rm s}}\sim 2\times 10^2$ d \textcolor{black}{\cite{1996ASPC..105..525A}}) orbited by quite distant circumbinary planet-like companion\footnote{For exoplanets, see \cite{exop1,exop2,exop3}.} ($P^{'}_{\rm b}\sim 3\times 10^4$ d).
For a discussion of the rate of change of the eccentricity in binaries hosting compact objects, see, e.g., \cite{Freire012}.

Our results are important also to accurately asses the overall uncertainty which can be obtained when constraints on $\Delta$ are inferred by comparing them with the corresponding  empirically determined quantities. Indeed,  taking into account only the accuracy with which an observable like, say, $de/dt$ can be determined is, in principle, not enough; a propagation of the errors affecting all the parameters \textcolor{black}{such as $g$ \cite{gala1,gala2,MW}},
and its orientation $\hat{\bds{g}}$, entering the corresponding theoretical prediction  must be done as well to correctly infer the total, systematic uncertainty in $\Delta$.

Our expressions are also useful in designing suitable SEP tests and in better interpreting present and future experiments, \textcolor{black}{not necessarily limited to binary pulsars}, especially when different competing theoretical mechanisms yielding SEP violations are considered. Indeed, we found that the ratio of the node and inclination SEP\textcolor{black}{-violating} precessions depends only on the inclination itself and on the pericenter.

\section*{Acknowledgments}
I thank P. Freire for useful correspondence and references. I am also grateful to an anonymous referee for her/his helpful comments.
\bibliography{SEPbib}{}

\end{document}